\documentclass[11pt]{article}

\usepackage{amsmath}
\usepackage{amsfonts}
\usepackage{amssymb}
\usepackage{graphicx}
\usepackage{supertabular}
\usepackage{setspace}

\RequirePackage{cite}%
\renewcommand{\citeleft}{\bgroup\normalfont[}%
\renewcommand{\citeright}{]\egroup}%

\newcommand{\n}{\noindent}
\usepackage{mathptmx}

\newcommand{\nn}{\nonumber}

\newcommand{\be}{\begin{equation}}
\newcommand{\ee}{\end{equation}}
\newcommand{\ba}{\begin{eqnarray}}
\newcommand{\ea}{\end{eqnarray}}
\newcommand{\bal}{\begin{align}}
\newcommand{\eal}{\end{align}}

\newcommand{\dd}{{\rm d}}

\newcommand{\bb}{\bibitem}

\begin{document}

\title{\leftline{\small Published in General Relativity and Gravitation (\texttt{www.springerlink.com})}
\leftline{\small M. Azreg-A\"{\i}nou, \textit{Gen. Relativ. Gravit.}, DOI 10.1007/s10714-012-1390-z}
\vskip0.9cm
Rotation and twist regular modes for trapped ghosts}
\author{Mustapha Azreg-A\"{\i}nou
\\
Ba\c{s}kent University, Department of Mathematics, Ba\u{g}l\i ca Campus, Ankara, Turkey}
\date{}

\maketitle

\begin{abstract}
A parameter-independent notion of stationary slow motion is formulated then applied to the case of stationary rotation of massless trapped ghosts. The excitations correspond to a rotation mode with angular momentum $J\neq 0$ and twist modes. It is found that the rotation mode, which has no parity, causes excess in the angular velocity of dragged distant coordinate frames in one sheet of the wormhole while in the other sheet the angular velocity of the ghosts is that of rotating stars: $2J/r^3$. As to the twist modes, which all have parity, they cause excess in the angular velocity of one of the throat's poles with respect to the other.

\vspace{3mm}

\n {\footnotesize\textbf{keywords.} Classical general relativity; Exact solutions; Self-gravitating systems}\\
\n {\footnotesize\textbf{PACS.} 04.20.-q; 04.20.Jb; 04.40.-b}

\vspace{-3mm} \n \line(1,0){431} % for no spacing
\end{abstract}

\section{Introduction}

We consider the action
\begin{equation}\label{1.1}
S=\int \dd ^{4}x \sqrt{|g|}\,[\mathcal{R}-\frac{1}{2}\,h(\phi) g^{\mu\nu}
\partial_{\mu}\phi\partial_{\nu}\phi +V(\phi)]\,,
\end{equation}
(where $\mathcal{R}^{\mu}{}_{\nu\rho\sigma}=
-\partial_{\sigma}\Gamma^{\mu}{}_{\nu\rho}+\cdots$
and the metric $g_{\mu\nu}$ has signature $+,-,-,-$), which
describes general relativity with a minimally coupled scalar field
$\phi$ as a source. We make all other conventions such that the Einstein equations take the form $G_{\mu}{}^{\nu}=T_{\mu}{}^{\nu}$ with a stress-energy tensor given by
\begin{equation}\label{1.2}
T_{\mu}{}^{\nu}=h\partial_{\mu}\phi\partial_{\rho}\phi g^{\rho\nu} - \frac{h}{2}\,\delta_{\mu}{}^{\nu}g^{\rho\sigma}\partial_{\rho}\phi\partial_{\sigma}\phi + \delta_{\mu}{}^{\nu}V\,.
\end{equation}
The metric-scalar field equations derived from~\eqref{1.1} read respectively
\begin{align}
\label{1.2a}&G_{\mu}{}^{\nu}=T_{\mu}{}^{\nu},\\
\label{1.2b}&\partial_{\mu}(\sqrt{|g|}hg^{\mu\nu}
\partial_{\nu}\phi)+\frac{\dd V}{\dd \phi}-\frac{1}{2}\frac{\dd h}{\dd \phi}g^{\mu\nu}
\partial_{\mu}\phi\partial_{\nu}\phi=0.
\end{align}

The function $h(\phi)$ ensures that the kinetic term in~\eqref{1.1}
has the correct sign ($h(\phi)>0$) far away from the source (in the weak-field region) and
has the wrong sign ($h(\phi)<0$) in the strong-field region located near the source,
which is assumed to be of phantom nature there~\cite{kmr,bs}. The purpose of introducing a non-constant factor $h$ is to be able to describe more plausible wormhole configurations in which the exotic part of the matter distribution, necessary for generating the throat, does not extend to accessible regions to observers, which are the weak-field regions, and remains confined near the throat. Thus violations of the energy conditions, particularly of the null energy condition (NEC), take place only in the strong-field region.

That such wormhole configurations exist, the so called ``trapped ghosts", where the field $\phi$ changes nature from phantom near the throat to normal away from it \textsl{without creating spacetime singularities}, has been proven in~\cite{bs}, the authors of which have shown that trapped ghosts are only possible with a non-vanishing potential $V(\phi)$. Moreover, Bronnikov and Sushkov have shown that the potential has both signs for any wormhole solution with two flat asymptotics. Explicitly, they have managed to construct regular, static, spherically symmetric massless trapped ghosts, where the latter are confined to the strong-field region ($h<0$), and have shown that the NEC is violated~\cite{bs} only in the region where the ghosts are trapped ($h<0$).

Some rotating classical wormholes are endowed with the property that some observer travelers may avoid the exotic matter~\cite{teo}. Some other properties remain to be discovered. From this point of view, it is interesting to discover the peculiar properties of rotating trapped ghosts, which may distinguish them from other rotating configurations. The aim of this work is to achieve that by constructing the rotation and twist regular counterparts of the trapped ghost found in~\cite{bs}, which are first reviewed in section 2. In section 3, we derive the most general master equation for stationary rotating configuration, define a notion of twist mode and generalize the existing parameter-dependent notions of slow rotation to a parameter-independent stationary slow motion notion. In section 4, we derive an exact solution for the rotation mode and series solutions for the twist modes and conclude.

%%%%%%%%%%%%%%%%%%%%%%%%%%%%%%%%%%%%%%%%%%%%%%%%%%%%%%%%%%%%%%%%%%%%%%%%%%%%%%%%%

%\setcounter{equation}{0}
\section{Static wormholes}

With a slight different notation than~\cite{bs}, we write the metric of a static, spherically symmetric configuration as
\begin{equation}\label{2.1}
    \dd s^2 = A(r)\,\dd t^2 - \frac{\dd r^2}{A(r)} - R^2(r)(\dd \theta^2 + \sin^2\theta \,\dd \varphi^2)\,.
\end{equation}
For a metric of the form~\eqref{2.1} the non-vanishing components of $G_{\mu}{}^{\nu}$ and $T_{\mu}{}^{\nu}$ along with the metric-scalar static field equations are given in the Appendix section.

Employing the inverse problem method, Bronnikov and Sushkov have chosen $R(r)$ and $\phi(r)$ as follows
\begin{align}
\label{2.2} & R(r)=\frac{r^2+\lambda a^2}{\sqrt{r^2+\lambda ^2 a^2}}\,,\qquad \lambda >2\,,\quad -\infty <r<+\infty\\
\label{2.3} & \phi(r)=\frac{2\phi_0}{\pi}\,\arctan\frac{r}{\lambda a}\,,\qquad \phi_0 = \frac{\pi a}{2}\,\sqrt{\frac{2(\lambda-2)}{\lambda}}\,,
\end{align}
then used the field equations (Eqs. 10 to 14 of~\cite{bs} or Eqs.~\eqref{a4} to~\eqref{a7}) to determine $A(r)$, $V(\phi)$ and $h(\phi)$ by
\begin{align}
 & A=\frac{3r^4+3\lambda (1+\lambda )a^2r^2+\lambda ^2(1+\lambda +\lambda
^2)a^4}{3 (r^2+\lambda  a^2)(r^2+\lambda ^2 a^2)}\,,\nn\\
 & V=\frac{\lambda ^2(\lambda -1)^2 a^4[-6r^4+\lambda (\lambda -5)a^2r^2 +\lambda ^3(\lambda +1)a^4]}{3(r^2+\lambda a^2)^2(r^2+\lambda^2a^2)^3}\,,\nn\\
\label{2.6} & h=\frac{(\lambda -2)r^2+\lambda ^2(1-2\lambda )a^2}{(\lambda -2)a^2(r^2+\lambda a^2)}\,,
\end{align}
where, for $-\phi_0<\phi<\phi_0$, $r=\lambda a\tan[\pi \phi/(2\phi_0)]$ may be substituted into the last two Eqs.~\eqref{2.6} to eliminate $r$. For $|\phi|\geq \phi_0$, one extends $V(\phi)$ and $h(\phi)$ by taking $V\equiv 0$ and $h\equiv 1/a^2$~\cite{bs}.

The functions~\eqref{2.2} to~\eqref{2.6} have all defined parities: except the scalar field $\phi(r)$ which is an odd function of $r$, the other functions are all even functions of $r$.

The solution~\eqref{2.2} to~\eqref{2.6} is a static, spherically symmetric, regular, massless and asymptotically flat wormhole with a throat of radius $a$, which is the minimum value of $R(r)$: $a\equiv R(r=0)$. The factor $h$ is positive for $|r|>r_0\equiv a\lambda \sqrt{(2\lambda -1)/(\lambda -2)}$ and negative for $|r|<r_0$. The function $r_0(\lambda)$ reaches its absolute minimum value at $\lambda_c=(13+\sqrt{105})/8\simeq 2.9$ with $r_0(\lambda_c)\simeq 6.7a$. Thus the exotic matter is distributed insight a radius of at least $6.7$ times the size of the throat $a$, which is an \textsl{arbitrary parameter}. Because of this property, the solution has been called a trapped ghost.

%%%%%%%%%%%%%%%%%%%%%%%%%%%%%%%%%%%%%%%%%%%%%%%%%%%%%%%%%%%%%%%%%%%%%%%%%%%%%%%%%

%\setcounter{equation}{0}
\section{Rotation and twist modes: General considerations}

The metric of a stationary and axially symmetric configuration, with two commuting Killing vectors $\partial_t$ (timelike) and $\partial_{\varphi}$ (spacelike), may be brought to the form~\cite{jbh1}
\begin{equation}\label{3.1}
\dd s^2 = \alpha(r,\theta)\dd t^2 - \beta(r,\theta)\dd r^2 - \gamma(r,\theta)\{\dd \theta^2 + \sin^2\theta [\dd \varphi - \Omega(r,\theta)\dd t]^2\},
\end{equation}

From now on, we use the ``algebraic" coordinates $(t,r,u,\varphi)$ where $u\equiv \cos\theta$. The metric and the fields $(\phi,V,h)$ are assumed to be functions of $(r,u)$:
\begin{equation}\label{3.2}
 \dd s^2 = \alpha\dd t^2 - \beta\dd r^2 - \gamma\bigg\{\frac{\dd u^2}{1-u^2} + (1-u^2) [\dd \varphi - \Omega\dd t]^2\bigg\}.
\end{equation}

For non-vanishing total angular momentum $J$, the function $\Omega(r,u)$, which is a local angular velocity of freely falling particles, admits \textsl{by definition} the following Taylor series as $r\to \infty$~\cite{lan,teo,p}
\begin{equation}\label{3.3}
    \Omega = 2J/r^3+\mathcal{O}(1/r^4)\,,
\end{equation}
and the corresponding solution is interpreted as rotating about its symmetry axis with angular momentum $J$~\cite{teo,p}.

\paragraph{\textbf{Definition.}} \textsl{For a stationary and axially symmetric solution with angular momentum $J$, each term proportional to $1/r^n$ and $n>3$ in the Taylor series in $1/r$ of the angular velocity $\Omega(r,u)$ corresponds to a twist mode of the solution about its symmetry axis}.

In the multi-pole expansion of $\Omega(r,u)$,
\begin{equation}\label{3.3a}
    \Omega(r,u)=\frac{2J}{r^3}+\sum_{n=4}^{\infty}\frac{Q(u)}{r^n}\,,
\end{equation}
the sign and magnitude of the coefficient $Q(u)$ is generally not constant and depends on $u$ (the angle $\theta$). Thus, the effect of each twist mode, acting separately, is that particles that fall freely from spatial infinity acquire different (in direction and magnitude) angular velocities as they reach the same surface $r=r_0$. As we shall see below this is possible for stationary and axially symmetric trapped ghosts; however, in a realistic wormhole configuration~\cite{khat} the twist modes may be ruled out physically.

%%%%%%%%%%%%%%%%%%%%%%%%%%%%%%%%%%%%%%%%%%%%%%%%%%%%%%%%%%%%%%%%%%%%%%%%%%%%%%%%%

%\setcounter{equation}{0}
\section{Slow rotation and twist mode wormholes}

Let $a_0$ be a rotation parameter. In our system of coordinates $(t,r,u,\varphi)$, which is adapted to the Killing vectors ($\partial_t,\partial_{\varphi}$), the two coordinates $(t,\varphi)$ are cyclic. This means that the metric~\eqref{3.2} remains invariant under translations back and forward along the orbits of ($\partial_t,\partial_{\varphi}$). This is a statement of invariance under reversal in the directions of both rotation and time and implies that the three metric components ($\alpha,\beta ,\gamma$) are even functions of $a_0$. As to $\Omega$, it depends on ${\rm sgn}\,(a_0)$ and so is an odd function of $a_0$. It is obvious that the fields ($\phi(r,u),V(r,u),h(r,u)$) do not depend on ${\rm sgn}\,(a_0)$ and so are even functions of $a_0$. Said otherwise, the values of the fields ($\phi(r,u),V(r,u),h(r,u)$) remain the same under reversal in the direction of rotation, which corresponds to the transformation: $a_0\to-a_0$~\cite{jbh1,sec}. Thus, to the first order in the rotation parameter $a_0$, which is the relevant order when studying the slow rotation of objects in order to determine their angular velocity\footnote{The second order approximation in $a_0$ is relevant to determine other physical properties of the rotating wormhole, like its new mass, and to investigate whether the NEC depends on rotation~\cite{sec}.}, the metric and the source admit the following expansions in terms of $a_0$~\cite{jbh1,sec}
\begin{align}\label{4.0}
&\alpha(r,u)=A(r)+\mathcal{O}(a_0^2),& & \beta(r,u)=\frac{1}{A(r)}+\mathcal{O}(a_0^2),&\nn\\ &\gamma(r,u)=R(r)^2+\mathcal{O}(a_0^2)\,,& & \Omega(r,u)=a_0\omega(r,u) +\mathcal{O}(a_0^3),\nn\\
&\phi(r,u)=\phi(r)+\mathcal{O}(a_0^2),& & V(r,u)=V(r)+\mathcal{O}(a_0^2),\nn\\
&h(r,u)=h(r)+\mathcal{O}(a_0^2),
\end{align}
where ($A(r),R(r),\phi(r),V(r),h(r)$) are the static values given by Eqs.~\eqref{2.2} to~\eqref{2.6} and $\omega(r,u)$ is to be fixed. So to this first order approximation there is only one function to determine, $\omega(r,u)$ (which is zero in the static case), and the other fields retain their static values. The metric and the source reduce to static values in the limit of zero rotation and spherical symmetry.

In order to determine $\omega(r,u)$ we need to solve the relevant equation. Since only $T_{\varphi}{}^{t}$ characterizes the rotation of matter distribution~\cite{teo}, the relevant equation is $G_{\varphi}{}^{t}=T_{\varphi}{}^{t}$. For stationary and axially symmetric configurations~\eqref{3.2}, the structure of $T_{\mu}{}^{\nu}$, as expressed in~\eqref{1.2}, is such that only the components ($T_{t}{}^{t},T_{r}{}^{r},T_{u}{}^{u},T_{\varphi}{}^{\varphi},T_{u}{}^{r}$) are non-zero. Thus, by the field equations $G_{\mu}{}^{\nu}=T_{\mu}{}^{\nu}$, $G_{\varphi}{}^{t}= 0$ since $T_{\varphi}{}^{t}\equiv 0$. The expression of $G_{\varphi}{}^{t}$ $(= R_{\varphi}{}^{t})$ is easily brought to the symmetrical form
\begin{equation}\label{3.4}
4(u^2-1)^3\beta \gamma^3\partial _r\Omega \partial _u\Omega G_{\varphi}{}^{t} =  \gamma \partial _r\Omega \partial _u\bigg[\frac{[(u^2-1)^2\beta \gamma \partial _u \Omega]^2}{\alpha \beta
}\bigg]-(u^2-1)\beta \partial _u\Omega \partial _r\bigg[\frac{[(u^2-1)\gamma^2\partial _r\Omega ]^2}{\alpha \beta}\bigg].
\end{equation}
Eq.~\eqref{3.4} is very useful for investigating stationary and axially symmetric configurations, as we shall see below, and will be used in future studies. Similar, however, special formulas have been derived and used in~\cite{jbh1,ks,hs}. The requirement $G_{\varphi}{}^{t}=0$ implies the vanishing of the r.h.s. in~\eqref{3.4}, which is highly nonlinear and in order to solve it we shall restrict ourselves to slow rotation and twist modes.

\subsection{\textbf{Slowly evolving stationary configurations}} We generalize the above considerations, which have led to~\eqref{4.0}, to the general case of stationary slow motion. A notion of slow rotation (of stars) in terms of the mass, (constant) angular velocity and radius of the mass distribution is given in~\cite{jbh1}. The definition has been extended to that of slowly rotating massive wormholes~\cite{ks} and is given in terms of only the throat radius and the throat equatorial angular velocity $\Omega(r=a,u=0)$. Since it does not depend on the mass of the wormhole, this definition applies to the rotation ($J\neq 0$) counterpart of the massless solution~\eqref{2.2} to~\eqref{2.6} but not to its twist modes, if considered separately, since as we shall see below, $\Omega(r,u=0)\equiv 0$ for some twist modes. We intend to give here a general, parameter-independent, definition of \textsl{slow evolution of stationary configurations}, which applies to all kind of \textsl{slow motion}. We assume that the static configuration admits two commuting Killing vectors, one timelike $\partial_t$ and the other spacelike $\partial_{x\,}$. In the system of coordinates $(t,x,y_i)$ adapted to $(\partial_t,\partial_x)$, where $y_i$ denote the remaining coordinates, the metric of the \textsl{static} configuration may be brought to a \textsl{diagonal matrix} $\pmb g_{\text{s}}(y_i)$~\cite{bh}.

\paragraph{\textbf{Definition.}} \textsl{Let $a_0$ be a motion parameter. We say that the physical configuration is in stationary slow motion in the ($+$ or $-$) direction of the orbits of $\partial_x$ if: 1) the metric} $\pmb g_{\text{d}}(y_i,a_0)$ \textsl{of the dynamic configuration (no longer diagonal) still admits $(\partial_t,\partial_x)$ as commuting Killing vectors;} 2) \textsl{The matrix} $\pmb g_{\text{d}}(y_i,a_0)$ \textsl{is analytic with respect to $a_0$\footnote{Each entry of it admits a Taylor expansion in powers of $a_0$.}; 3) The diagonal part of} $\pmb g_{\text{d}}(y_i,a_0)$\textsl{, denoted by} ${\rm diag}(\pmb g_{\text{d}}(y_i,a_0))$\textsl{, is an even matrix function of $a_0$; 4) The off-diagonal part of} $\pmb g_{\text{d}}(y_i,a_0)$\textsl{, that is} $\pmb g_{\text{d}}(y_i,a_0)-{\rm diag}(\pmb g_{\text{d}}(y_i,a_0))$\textsl{, is an odd matrix function of $a_0$, which goes to zero as $a_0\to 0$; 5) The max norm\footnote{The max norm of a matrix is the largest entry of it in absolute value.} of the matrix} ${\rm diag}(\pmb g_{\text{d}}(y_i,a_0))-\pmb g_{\text{s}}(y_i)$ \textsl{is much smaller than unity.}

It follows that ${\rm diag}(\pmb g_{\text{d}}(y_i,a_0))$ and $\pmb g_{\text{d}}(y_i,a_0)-{\rm diag}(\pmb g_{\text{d}}(y_i,a_0))$ are expanded as, respectively
\begin{align}
\label{4.1} &{\rm diag}(\pmb g_{\text{d}}(y_i,a_0))=  \pmb g_{\text{s}}(y_i,a_0)+a_0^2 \pmb P(y_i)+\mathcal{O}(a_0^4),\\
\label{4.2} &\pmb g_{\text{d}}(y_i,a_0)-{\rm diag}(\pmb g_{\text{d}}(y_i,a_0))= a_0 \pmb \omega(y_i)+\mathcal{O}(a_0^3)\,,
\end{align}
where the two matrices $(\pmb P(y_i),\pmb \omega(y_i))$ are assumed to be \textsl{regular} on the whole ranges of $y_i$. Notice that a master equation providing an expression for $G_{x}{}^{t}$, similar to~\eqref{3.4}, may easily be derived.

If we restrict ourselves to the first order in $a_0$ and apply the above considerations to the stationary metric~\eqref{3.2}, where $\partial_x=\partial_{\varphi}$, $y_i=(r,u)$ and $a_0$ is a rotation parameter and using~\eqref{2.1} (the static metric), we are led to~\eqref{4.0}, where $\omega$ is\footnote{This should be denoted by $\omega_{t\varphi}$ but we have omitted the indices $t,\,\varphi$.} the unique entry of the matrix $\pmb \omega$. Thus to the first order in $a_0$, which is the relevant order to our investigation, we approximate the functions ($\alpha,\beta,\gamma$) by their zero-order static values ($A,1/A,R^2$) and the same applies to ($\phi(u,r),V(u,r),h(u,r)$). To this order of approximation, all the field equations are automatically satisfied except $G_{\varphi}{}^{t}=0$, the r.h.s of~\eqref{3.4}, which becomes \textsl{linear} in $\omega(r,u)$
\begin{equation}\label{4.5}
 R^2\,\partial_u[(u^2-1)^2\,\partial_u\omega]=(u^2-1)A\,\partial_r(R^4\,\partial_r\omega)\,.
\end{equation}

\subsection{\textbf{The solutions: regular modes}} The general solution to the \textsl{linear} partial differential equation~\eqref{4.5} may be constructed upon separating the variables. Setting $\omega(r,u)=f(r)g(u)$ and $\eta={\rm constant}$, we obtain
\begin{equation}\label{4.6}
    \frac{A}{R^2}\frac{\partial_r(R^4\partial_rf)}{f}=
    \frac{1}{u^2-1}\frac{\partial_u[(u^2-1)^2\partial_ug]}{g}=\eta.
\end{equation}
As $r\to\infty$, the functions $(A,R,f)$ approach the limits
\begin{equation}\label{n1}
A\to 1\,,\;R\to r\,,\;f\to r^{-m}\qquad \text{($m\geq 3$ by~\eqref{3.3})}\,.
\end{equation}
Substituting these limits into~\eqref{4.6} we obtain $\eta=m(m-3)$.

\subsubsection{\textbf{The} $\pmb u$\textbf{-equation}} The behavior of $g_m(u)$ is governed by the equation
\begin{equation}\label{4.7}
 \partial_u[(u^2-1)^2\partial_ug_m]=m(m-3)(u^2-1)g_m,\quad m\geq 3,
\end{equation}
whose solution is a linear combination of the Gauss hypergeometric function $_2F_1\big((3-m)/2,\,m/2;\,1/2;\,u^2\big)$ and $_2F_1\big((4-m)/2,\,(m+1)/2;\,3/2;\,u^2\big)u$ and it diverges at the poles $u=1$ or $u=-1$ ($\theta=0$ or $\theta=\pi$). From these linearly independent solutions one may choose the everywhere regular solutions as follows. If $m\geq 3$ is an odd integer, then $g_m(u)= {_2}F_1\big((3-m)/2,\,m/2;\,1/2;\,u^2\big)$ is an even polynomial in $u$ of degree $m-3$; If $m\geq 4$ is an even integer, then $g_m(u)= {_2}F_1\big((4-m)/2,\,(m+1)/2;\,3/2;\,u^2\big)u$ is an odd polynomial in $u$ of degree $m-3$:
\begin{align}
\label{4.8}& m\text{ (integer) }\equiv 2n+1\geq 3\Rightarrow \nn\\
&g_m(u)= {_2}F_1\big(1-n,1/2+n;1/2;u^2\big):\text{ even polynomial},\\
\label{4.9}& m\text{ (integer) }\equiv 2n\geq 4\Rightarrow \nn\\
&g_m(u)= {_2}F_1\big(2-n,1/2+n;3/2;u^2\big)u:\text{ odd polynomial}.
\end{align}
The first three modes are
\begin{equation}\label{10}
m= 3\Rightarrow g_3(u)=1;\;\;m= 4\Rightarrow g_4(u)= u;\;\;m= 5\Rightarrow g_5(u)= -5u^2+1.
\end{equation}
For the mode $m=4$, two test particles falling freely from $r=\infty$ to the points $(r,u)$ and $(r,-u)$ will have opposite local angular velocities, $a_0uf(r)$ and $-a_0uf(r)$ while for the mode $m=5$, the same particles reaching $(r,u_1)$ and $(r,u_2)$, will rotate in different directions if, for instance, $\sqrt{1/5}<u_1<1$ and $0<u_2<\sqrt{1/5}\,$.

\subsubsection{\textbf{The} $\pmb r$\textbf{-equation}} The behavior of $f_m(r)$ is governed by the following equation where $m$ is integer
\begin{equation}\label{4.11}
 \partial_r(R^4\partial_rf_m)=m(m-3)(R^2/A)f_m,\quad m\geq 3.
\end{equation}
\paragraph{\textbf{(i) Case} $\pmb{m=3}$.} The regular solution for the rotation mode $m=3$, with angular momentum $J$, is given by
\begin{multline}\label{4.12}
    f_3(r)=+\frac{3(J/a_0)(5 \lambda ^2+2\lambda +1)}{8a^3\lambda ^{3/2}}\bigg[\frac{\pi }{2}-\arctan\bigg(\frac{r}{a\sqrt{\lambda }}\bigg)\bigg]\\
    -\frac{(J/a_0)}{8} \bigg[\frac{8a^2(\lambda -1)^2\text{$\lambda $}r}{(r^2+a^2\lambda )^3}+\frac{2(5 \lambda ^2+2\lambda -7)r}{(r^2+a^2\lambda)^2}+\frac{3(5 \lambda ^2+2\lambda +1)r}{a^2\lambda (r^2+a^2\lambda )}\bigg].
\end{multline}

Now, we introduce a new parameter, $\Omega_{\rm th}$, the angular velocity of the throat defined by
\begin{equation}\label{p1}
\Omega_{\rm th}\equiv \Omega(r=0,\theta=\pi/2)=a_0\omega(r=0,u=0)\,,
\end{equation}
in terms of which the slow motion hypothesis implies
\begin{equation}\label{p2}
    \Omega_{\rm th}a\ll c(\equiv 1)\,.
\end{equation}

As will be established below, we will construct a solution such that
\begin{equation}\label{p3}
\omega(r=0,u=0)\equiv \sum_{m=3}f_m(r=0)g_m(u=0)=f_3(r=0)=\frac{3\pi}{16}\frac{J}{a_0}\frac{(5 \lambda ^2+2\lambda +1)}{a^3\lambda ^{3/2}},
\end{equation}
(where we have added the $m$-mode solutions to the linear equation~\eqref{4.5}). Thus, using~\eqref{p1} we obtain
\begin{equation}\label{p4}
    J=\frac{16}{3\pi}\frac{a^3\lambda ^{3/2}\Omega_{\rm th}}{(5 \lambda ^2+2\lambda +1)}\,,
\end{equation}
and the condition~\eqref{p2} becomes
\begin{equation}\label{p5}
    J\ll \frac{16}{3\pi}\frac{a^2\lambda ^{3/2}c}{(5 \lambda ^2+2\lambda +1)}\,.
\end{equation}

Eq.~\eqref{4.12} implies that as $r\to\infty$, $f_3\to 2(J/a_0)/r^3$ but as $r\to -\infty$, $f_3\to f_3(-\infty)+2(J/a_0)/r^3$. Thus, distant coordinate frames at $r=+\infty$ are not dragged while those at $r=-\infty$ rotate with an angular velocity of
\begin{equation*}
a_0\omega_3(-\infty)=a_0f_3(-\infty)=\frac{\pi(5 \lambda ^2+2\lambda +1)J}{8a^3\lambda ^{3/2}}=\frac{2}{3}\Omega_{\rm th}\,.
\end{equation*}
The same effect has been observed for slowly rotating Ellis wormholes~\cite{ks}.

\begin{figure}[h]
\centering
  \includegraphics[width=0.3\textwidth]{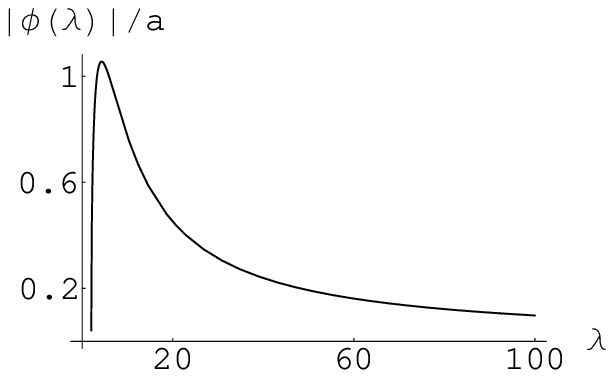} \includegraphics[width=0.4\textwidth]{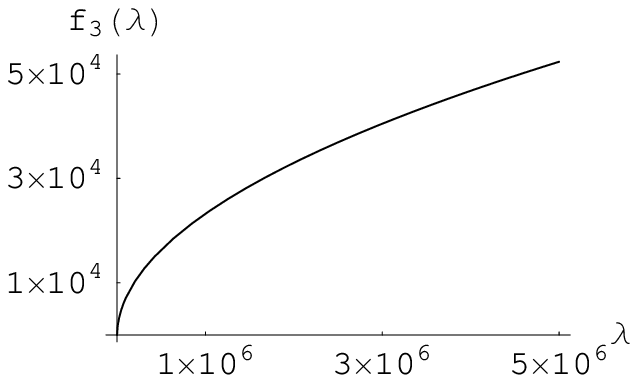}\\
  \caption{\footnotesize{In both plots $r/a=7$. (Left) The scalar field. (Right) The angular velocity in units of $J/(8a_0a^3)$.}}\label{Fig1}
\end{figure}

\begin{figure}[h]
\centering
  \includegraphics[width=0.5\textwidth]{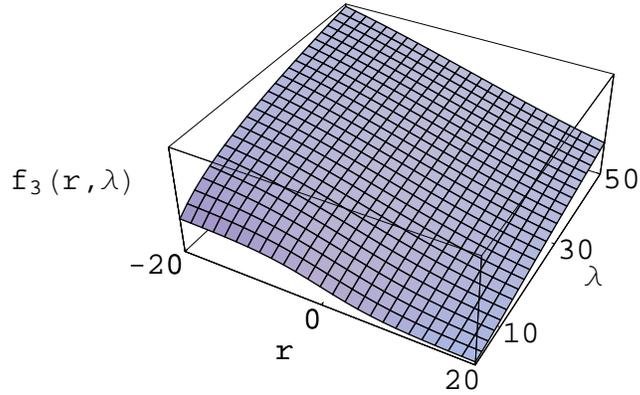}\\
  \caption{\footnotesize{The angular velocity in units of $J/(8a_0)$ for $a=10$.}}\label{Fig2}
\end{figure}

The damping effect of the scalar field $\phi$ was discussed in~\cite{man} for the case of zero-mass scalar field. For the case of rotating trapped ghosts, this effect is better illustrated in Figure~\ref{Fig1}. For fixed $r$ and $a$ (precisely for fixed $r/a$), the scalar field in absolute value, $|\phi(\lambda)|$, is a decreasing function of $\lambda$, except for the small values of $\lambda$: $2<\lambda <\lambda_{\rm max}$, where $\lambda_{\rm max}$ is such that $\frac{\dd \phi}{\dd \lambda}(\lambda_{\rm max})=0$. While $f_3(\lambda)$ increases as $\lambda$ does. The damping effect of the scalar field is observed in both asymptotically flat regions $r\to \pm\infty$ as shown in Figure~\ref{Fig2}.

\paragraph{\textbf{(ii) Case} $\pmb{m>3}$.} For the twist modes $m>3$, there are no solutions in closed forms to~\eqref{4.11}. However, if we bring the latter to the form
\begin{equation}\label{f1}
    r^2\partial_{rr}f_m + rp(r)\partial_{r}f_m + q(r)f_m=0\,,\quad (m>3)
\end{equation}
with
\begin{equation}\label{f2}
    p(r)\equiv \frac{4r}{R(r)}\partial_{r}R(r)\,,\quad q(r)\equiv -\frac{m(m-3)r^2}{A(r)R^2(r)}\,,
\end{equation}
we can solve it by the method of Frobenius.

The coefficients $p(r)$, $q(r)$ have even power series expansions vanishing at $r=0$, which is an ordinary point of~\eqref{f1}
\begin{equation}\label{f3}
    p(r)=\sum_{i=1}^{\infty}p_{2i}r^{2i}\,,\quad q(r)=\sum_{i=1}^{\infty}q_{2i}(m)r^{2i}\,,
\end{equation}
with
\begin{equation*}
p_2=-\frac{4(1-2\lambda)}{\lambda^2a^2},\;q_2=-\frac{3m(m-3)\lambda^2}{a^2(\lambda^3+\lambda^2+\lambda)},\cdots\,,
\end{equation*}

We look for solutions to~\eqref{f1} of the form
\begin{equation}\label{f4}
    f_m(r)=r^s\sum_{k=0}^{\infty}c_k(m)r^k\,,
\end{equation}
leading, using~\eqref{f3} and~\eqref{f4} in~\eqref{f1}, to $s=s_1\equiv 1$, $s=s_2\equiv 0$
which solve the indicial equation: $s(s-1)c_0=0$ where $c_0$ is an arbitrary constant. The general solution to~\eqref{f4} splits into a sum of linearly independent even and odd solutions:
\begin{equation}\label{f6}
    f_m(r)=\sum_{k=0}c_{2k}(m)r^{2k} + \sum_{k=0}c_{2k+1}(m)r^{2k+1}\,,
\end{equation}
where $c_0$, $c_1$ remain arbitrary constants. Furthermore, due to the properties~\eqref{f3}, $c_{2k}$ depend on $c_0$ but not on $c_1$ and $c_{2k+1}$ depend on $c_1$ but not on $c_0$ according to the recurrence formulas:
\begin{align*}
& 2k(2k-1)c_{2k}=-\sum_{i=0}^{k-1}[2ip_{2(k-i)}+q_{2(k-i)}]c_{2i},\\
& 2k(2k+1)c_{2k+1}=-\sum_{i=0}^{k-1}[(2i+1)p_{2(k-i)}+q_{2(k-i)}]c_{2i+1}.
\end{align*}
For instance
\begin{align*}
& \frac{c_2}{c_0}=-\frac{q_2}{2}\,,\;\frac{c_4}{c_0}=\frac{2p_2q_2+q_2^2-2q_4}{24}\,,\\
& \frac{c_3}{c_1}=-\frac{p_2+q_2}{6}\,,\;\frac{c_5}{c_1}=\frac{3p_2^2+q_2^2+4p_2q_2-6(p_4+q_4)}{120}\,.
\end{align*}

The general asymptotic solution to~\eqref{f1} as $r\to \pm\infty$ is derived setting $\rho=1/r$ in terms of which~\eqref{f1} reads
\begin{equation}\label{pp1}
    \rho^2\partial_{\rho\rho}f_m + \rho\Big[2+4\Big(\frac{\rho}{R}\Big)\Big]\partial_{\rho}f_m
    + \frac{m(3-m)}{\rho^2R^2A}f_m=0\,.
\end{equation}
$\rho=0$ is a regular singular point and solutions to the indicial equation are $s=s_1\equiv m$, $s=s_2\equiv 3-m$ so that the general solution to~\eqref{pp1} reads\footnote{The roots $s_1$, $s_2$ are such that $s_1-s_2=2m-3$ is a positive integer, this corresponds to an exceptional case, as $\rho=0$ is a regular singular point, and may lead to a general solution to~\eqref{pp1} including a term in $\ln \rho$. However, since the coefficients in~\eqref{pp1} are all even functions of $\rho$, we verify that this leads to the so called ``false exception"~\cite{DE} where the general solution to~\eqref{pp1} does not include a term in $\ln \rho$.} (with $m>3$)
\begin{equation}\label{pp2}
    f_m = \rho^{3-m}\sum_{k=0}\kappa_{2k}(m) \rho^{2k} + \rho^m\sum_{k=0}\kappa_{2k+1}(m) \rho^{2k}\,,
\end{equation}
where $\kappa_0$, $\kappa_1$ are arbitrary constants. Furthermore, $\kappa_{2k}$ depend on $\kappa_0$ but not on $\kappa_1$ and $\kappa_{2k+1}$ depend on $\kappa_1$ but not on $\kappa_0$.

Now, since we look for regular solutions, we have to drop the first term in~\eqref{pp2} which diverges as $\rho^{3-m}$. The remaining term in~\eqref{pp2} is an odd or even function of $\rho$ (or $r$) if $m$ is odd or even, respectively, so that this term always corresponds to one of the two linearly independent parity-defined solutions in~\eqref{f6}. Thus, upon choosing $c_0=0$ or $c_1=0$ in~\eqref{f6}, we obtain an \textsl{everywhere} regular odd solution of $r$ if $m$ is odd or regular even solution of $r$ if $m$ is even, respectively, of the form
\begin{equation}\label{pp3}
m:\text{odd,}\,f_m(r)=\left\{
  \begin{array}{ll}
    \sum_{k=0}c_{2k+1}(m)r^{2k+1}, & \hbox{$r\to 0$;} \\
    \sum_{k=0}\dfrac{\kappa_{2k+1}(m)}{r^{m+2k}}, & \hbox{$r\to \pm \infty$,}
  \end{array}
\right.
\end{equation}
\begin{equation}\label{pp4}
m:\text{even,}\,f_m(r)=\left\{
  \begin{array}{ll}
    \sum_{k=0}c_{2k}(m)r^{2k}, & \hbox{$r\to 0$;} \\
    \sum_{k=0}\dfrac{\kappa_{2k+1}(m)}{r^{m+2k}}, & \hbox{$r\to \pm \infty$.}
  \end{array}
\right.
\end{equation}
Since $f_m(r)$ has a defined parity, its behavior at $-\infty$ is that at $+\infty$ with the appropriate sign: as $r\to -\infty$, $f_m(r)\to (-1)^{m}\kappa_{1}(m)/|r|^m$.

From equations~\eqref{4.9} and \eqref{pp3} we have $g_m(u=0)=0$ if $m$ is even and $f_m(r=0)=0$ if $m$ is odd, which justifies~\eqref{p3}.

\section{The angular velocity} The total angular velocity is $a_0\sum_{m=3}f_mg_m$, keeping only the two first terms we have
\begin{equation}\label{4.16}
    \Omega(r,\theta)=a_0f_3(r)+a_0\cos\theta f_4(r)+\cdots\,.
\end{equation}
At large distances from the exotic matter, the rotation mode ($m=3$) in~\eqref{4.16} is predominant and $\Omega(r,\theta)$ has the same sign on a ``sphere" $r=r_0\to\pm\infty$. Near the throat $r=0$,~\eqref{4.16} becomes
\begin{equation}\label{4.17}
    \Omega(r=0,\theta)=\Omega_{\rm th}+(\Omega_{\rm ex}/2)\cos\theta \,,
\end{equation}
where $\Omega_{\rm ex}\equiv 2a_0c_0(4)$ measures the excess angular velocity at the pole $(r,\theta)=(0,\pi)$ of the throat with respect to the pole $(0,-\pi)$. The twist term in~\eqref{4.17}, $(\Omega_{\rm ex}/2)\cos\theta$, causes falling particles in the region $\pi/2<\theta<\pi$ to slow down with respect to particles falling in the region $0<\theta<\pi/2$. The constants $\Omega_{\rm th}$ and $\Omega_{\rm ex}$ are independent; however, in a realistic wormhole configuration~\cite{khat} $\Omega_{\rm ex}$, which depends on the rotation parameter $a_0$, is assumed to be small compared to $\Omega_{\rm th}$.

\section{Conclusion}

The trapped exotic matter, responsible for maintaining the throat of the ghost, extends over 6.7 times the size of the throat.

In a stationary slow motion the diagonal part of the dynamic metric is an even function of the motion parameter and the remaining components of the metric are odd functions of it.

For slowly rotating trapped ghosts, the dragging effect vanishes as $r\to\infty$, but as $r\to-\infty$ the effect causes the angular velocity of distant coordinate frames to approach the two thirds of that of the throat. This excess angular velocity is due to the rotation mode. Due to damping effects, the dragging effect intensifies for low scalar field amplitudes.

Apart from the radial rotation mode, $f_3$, which has no parity, all radial twist modes, as well all angular rotation and twits modes, have defined parities. Due to this property, only the rotation mode contributes to the angular velocity of the throat; however, the twist modes contribute to the excess angular velocity.

It is worth mentioning that the application of the Newman-Janis algorithm~\cite{nj,kr,arg} to the static metric~\eqref{2.1} does never lead to a stationary metric of the form~\eqref{3.1}~\cite{arg,az2} whatever the type of complexification envisaged~\cite{kr}. The metric derived by the Newman-Janis trick, being endowed with rotation, however, cannot be brought to the Kerr-form~\eqref{3.1} by a coordinate transformation~\cite{az2}. This observation limits the use of the complexification trick, which is likely related to the nonlinearity of the source~\cite{arg}.

\section*{Appendix}
\appendix
\def\theequation{A.\arabic{equation}}
\setcounter{equation}{0}
In terms of $(t,r,u,\varphi)$ where $u=\cos\theta$, the static, spherically symmetric metric~\eqref{2.1} takes the form
\begin{equation}\label{a1}
    \dd s^2= A(r)\dd t^2 - \frac{\dd r^2}{A(r)} - R^2(r)\bigg[\frac{\dd u^2}{1-u^2} + (1-u^2) \dd \varphi^2\bigg].
\end{equation}

From now on we use the notation $'=\dd /\dd r$. For the static, spherically symmetric metric~\eqref{a1} the non-vanishing $G_{\mu}{}^{\nu}$ and $T_{\mu}{}^{\nu}$ read respectively
\begin{align*}
& G_{t}{}^{t}=\frac{1-A'RR'-A({R'}^2+2RR'')}{R^2}\,,\\
& G_{r}{}^{r}=\frac{1-A'RR'-A{R'}^2}{R^2}\,,\\
& G_{u}{}^{u}=G_{\varphi}{}^{\varphi}=-\frac{2A'R'+2AR''+A''R}{2R}\,,\\
& T_{t}{}^{t}=T_{u}{}^{u}=T_{\varphi}{}^{\varphi}=V+\frac{1}{2}Ah{\phi'}^2\,,\\
& T_{r}{}^{r}=V-\frac{1}{2}Ah{\phi'}^2\,.
\end{align*}
Thus the set of equations~\eqref{1.2a} are
\begin{align}\label{a3}
& G_{t}{}^{t}=T_{t}{}^{t}\Rightarrow\frac{1-A'RR'-A({R'}^2+2RR'')}{R^2}=V+\frac{1}{2}Ah{\phi'}^2\,,\nn\\
& G_{r}{}^{r}=T_{r}{}^{r}\Rightarrow\frac{1-A'RR'-A{R'}^2}{R^2}=V-\frac{1}{2}Ah{\phi'}^2\,,\\
& G_{u}{}^{u}=T_{u}{}^{u}\Rightarrow-\frac{2A'R'+2AR''+A''R}{2R}=V+\frac{1}{2}Ah{\phi'}^2\,.\nn
\end{align}
These equations lead to simplified forms if we combine them as follows. $G_{r}{}^{r}-G_{t}{}^{t}=T_{r}{}^{r}-T_{t}{}^{t}$ leads to
\begin{equation}\label{a4}
    2R''/R=-h{\phi'}^2\,,
\end{equation}
and $G_{r}{}^{r}+G_{u}{}^{u}=T_{r}{}^{r}+T_{u}{}^{u}$, using~\eqref{a4}, leads to
\begin{equation}\label{a5}
    A(R^2)''-A''R^2=2\,,
\end{equation}
and finally $G_{r}{}^{r}+G_{t}{}^{t}=T_{r}{}^{r}+T_{t}{}^{t}$, using~\eqref{a5}, leads to
\begin{equation}\label{a6}
    (A'R^2)'=-2R^2V\,.
\end{equation}
The equation~\eqref{1.2b} reads
\begin{equation}\label{a7}
    (AR^2h\phi')'-\frac{1}{2}AR^2h'\phi'=R^2\frac{\dd V}{\dd \phi}\,.
\end{equation}

\end{document}